\documentclass[pra,amsmath,amssymb, twocolumn, showpacs]{revtex4}

\usepackage{dcolumn}
\usepackage[dvips]{graphicx}

\usepackage{epsfig}

\begin{document}

\title{Algebraic characterization of $X$-states in quantum information}

\author{A.\ R.\  P. Rau$^{*}$}
\affiliation{Department of Physics and Astronomy, Louisiana State University,
Baton Rouge, Louisiana 70803-4001}


\begin{abstract}

A class of two-qubit states called $X$-states are increasingly being used to discuss entanglement and other quantum correlations in the field of quantum information. Maximally entangled Bell states and ``Werner" states are subsets of them. Apart from being so named because their density matrix looks like the letter X, there is not as yet any characterization of them. The $su(2) \times su(2) \times u(1)$ subalgebra of the full $su(4)$ algebra of two qubits is pointed out as the underlying invariance of this class of states. $X$-states are a seven-parameter family associated with this subalgebra of seven operators. This recognition provides a route to preparing such states and also a convenient algebraic procedure for analytically calculating their properties. At the same time, it points to other groups of seven-parameter states that, while not at first sight appearing similar, are also invariant under the same subalgebra. And it opens the way to analyzing invariant states of other subalgebras in bipartite systems.
    
\end{abstract}


\maketitle

\section{Introduction}

Increasingly in the field of quantum information, aspects of entanglement \cite{ref1}, and of other quantum correlations such as, for instance, ``quantum discord" \cite{ref2}, between two qubits have been described for a class of pure and mixed states that have come to be called ``$X$-states" \cite{ref3}. Although their use goes back further \cite{ref4}, they were so named in \cite{ref3} because of the visual appearance of the density matrix, that it looks like the letter in the alphabet:

\begin{equation}
\rho =  \left( 
\begin{array}{cccc}
\rho_{11} & 0 & 0 & \rho_{14} \\ 
0 & \rho_{22} & \rho_{23} & 0 \\ 
0 & \rho_{32} & \rho_{33} & 0 \\
\rho_{41} & 0 & 0 & \rho_{44}
\end{array}
\right).      
\label{eqn1}
\end{equation}

Non-zero entries occur only along the diagonal and anti-diagonal. Many calculations of entanglement and other properties \cite{ref4,ref5}, and their evolution under unitary or dissipative processes \cite{ref6}, can be carried out analytically for such states which make them appealing objects for study. Many other states of interest, such as the maximally entangled Bell states \cite{ref1} and ``Werner" states \cite{ref7}, are a sub-class of $X$-states, lending further importance to their study. 

Yet, no firmer definition has been given of what makes a pure or mixed system an $X$-state. This Letter provides such a definition in terms of their invariance properties, that a particular symmetry group or algebra underlies them. Such an identification of an underlying symmetry helps to explain the analytical results while at the same time providing a well defined procedure for their preparation. Recognizing the symmetry also makes computations involving such states, such as unitary operations on them or evaluating concurrence or other measures of entanglement, straightforward and easily tractable. And, finally, the symmetry also opens the way for constructing other density matrices which may not visually appear as X but are nevertheless similar, states of a different rendering of the same algebraic symmetry. Since they differ in entanglement and separability considerations, they may prove useful for study.
 
\section{The subalgebra of $X$-states}
     
Positivity and other standard requirements of any density matrix make the $X$-states shown in Eq.~(\ref{eqn1}) a seven-parameter family. The diagonal elements of the density matrix are real so that, along with the trace being fixed at 1, three real parameters describe those diagonal entries. Hermiticity to guarantee real eigenvalues reduces the off-diagonal entries to two complex (say $\rho_{14}$ and $\rho_{23}$, with $\rho_{41}$ and $\rho_{32}$ their respective complex conjugates) or four real parameters for the total of seven real parameters.

The full two qubit system has the symmetry of the SU(4) group and its algebra $su(4)$. Fifteen operators, most conveniently rendered as fifteen linearly independent 4 $\times$ 4 matrices or as Pauli spinors/matrices of the two spins, together with the unit matrix, provide a complete description of the general system. There are, however, several subalgebras of $su(4)$. A series of recent papers have provided a geometrical description of their states and operators \cite{ref8,ref9,ref10,ref11}. In particular, one subalgebra, $su(2) \times su(2) \times u(1)$, of seven operators or matrices occurs in many physical systems in quantum optics and quantum information \cite{ref8,ref9}. This Letter presents them as the invariance set of the $X$-states.

Inspection of the explicit 4 $\times$ 4 matrices in a standard basis for two spins, $(|\uparrow \uparrow 
\rangle, |\uparrow \downarrow \rangle, |\downarrow \uparrow \rangle, 
|\downarrow \downarrow \rangle)$, is instructive \cite{ref8,ref9,ref12,ref13} and points immediately to sets of seven of them with the same structure of eight zeroes in the same positions as in Eq.~(\ref{eqn1}). That is, these are operators that do not mix the 1-4 and 2-3 subspaces of the density matrix. Combined with the observation that such a set of seven matrices closes under multiplication, it is immediate that they will carry $X$-states into each other, that they preserve the X structure. For this purpose, both the Lie algebra aspect that the seven operators close under commutation and their Clifford algebraic structure that they close under multiplication are important. Indeed, explicit rendering of the fifteen operators in terms of two Pauli spinors called $\vec{\sigma}$ and $\vec{\tau}$, together with the familiar multiplication rule $\sigma_i \sigma_j = \delta_{ij} + i \epsilon_{ijk} \sigma_k, i, j, k = 1-3$, where $\epsilon_{ijk}$ is the completely antisymmetric symbol and repeated indices are summed, is very useful for operations with them. 

There are many such sets of seven operators/matrices constituting the $su(2) \times su(2) \times u(1)$ subalgebra \cite{ref8,ref9,ref11}. In each of them, one operator, the $u(1)$ element, commutes with all six of the others which themselves can be further subdivided as shown in \cite{ref8} into two sets of ``pseudospins", two sets of three which obey commutation relations of angular momentum within each set while all three of one set commute with all three of the other. Any one of the fifteen operators can serve as the commuting element because, as shown in a table in \cite{ref9}, each row has six zeroes so that each identifies such a $su(2) \times su(2) \times u(1)$ set. There are, therefore, fifteen non-equivalent such subalgebras. 

We will designate such a set by $\{X_i\}, i=1, 2, \ldots, 7$, with $X_1$ the commuting element. One such is $(X_1 =\sigma_z \tau_z, X_2=\sigma_y \tau_x, X_3=\tau_z, X_4=-\sigma_y\tau_y, X_5=\sigma_x\tau_y, X_6=\sigma_z, X_7=\sigma_x\tau_x)$. This is the same set that occurs in the CNOT quantum logic gate constructed out of two Josephson junctions and was extensively studied in that context \cite{ref8}. It was also pointed out that it occurs in nuclear magnetic resonance when each spin is in an external magnetic field in the $z$-direction while being coupled to each other through scalar coupling $\vec{\sigma} \cdot \vec{\tau}$ and ``cross-coherences" $\sigma_x\tau_y$ and $\sigma_y\tau_x$. But a different choice for the commuting element $X_1$ gives another such subalgebra, and we will return to this in section IV. Each $X_i$ is traceless, Hermitian, and unitary, and its square is unity so that the eigenvalues are ($\pm 1, \pm 1$).

With any such set, $\{X_i\}$, the density matrix that remains invariant under their operations can be rendered as a linear superposition of them,

\begin{equation}
\rho =(I + \Sigma_i g_i X_i)/4,
\label{eqn2}
\end{equation}
in analogy to that for a single spin, $(I+\Sigma_i g_i \sigma_i)/2$. The seven real coefficients $g_i$ in Eq.~(\ref{eqn2}) parametrize $X$-states and are equivalent to the seven parameters in the density matrix in Eq.~(\ref{eqn1}):

\begin{eqnarray}
g_1 & = & (\rho_{11} +\rho_{44}) -(\rho_{22} +\rho_{33}), \nonumber \\
g_2 & = & 2i(\rho_{14}-\rho_{41}+\rho_{32}-\rho_{23}), \nonumber \\
g_3 & = & (\rho_{11} -\rho_{44}) -(\rho_{22} -\rho_{33}), \nonumber \\
g_4 & = & 2(\rho_{14}+\rho_{41}-\rho_{32}-\rho_{23}), \nonumber \\ 
g_5 & = & 2i(\rho_{14}-\rho_{41}-\rho_{32}+\rho_{23}), \nonumber \\
g_6 & = & (\rho_{11} -\rho_{44}) +(\rho_{22} -\rho_{33}), \nonumber \\
g_7 & = & 2(\rho_{14}+\rho_{41}+\rho_{32}+\rho_{23}). 
\label{eqn3}
\end{eqnarray}

The algebra of the seven $\{X_i\}$ is most conveniently captured by Fig. 1 as has recently been pointed out \cite{ref11}. This figure occurs in projective geometry as the ``Fano Plane" \cite{ref14} and also is used to represent the multiplication table for octonions \cite{ref15}. Arranging the seven operators at the vertices, mid-points of sides and in-center of an equilateral triangle, the seven lines shown (including the inscribed circle) each carry three points, providing the multiplication rule for those $\{X_i\}$. The notation of arrows is also borrowed from octonions except that unlike them which have all seven lines arrowed, the three internal verticals are not in Fig. 1. On those lines, all three operators mutually commute, so that the product of two gives the third regardless of order. On the four arrowed lines, the operators mutually anticommute so that the product of two gives $(\pm i)$ times the third, with plus (minus) signs along (against) the sense of the arrow. For this purpose, each line is regarded as a closed loop with a continuously circulating arrow. The central element commutes with all six of the others. For each of those, there is one ``conjugate" element with which it commutes and four with which it anticommutes. All of this can be read off by merely glancing at Fig. 1 and will provide simple rules for their manipulation in the next section.

\begin{figure}
\vspace{-.5in}
\scalebox{1.6}{\includegraphics[width=2.in]{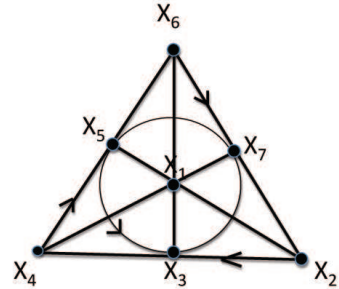}}
\vspace{-.4in}
\caption{The multiplication diagram for the seven operators that underlie $X$-states. Resembling the Fano Plane and the multiplication diagram for octonions, each operator stands on three lines, and each line, including the inscribed circle, has on it three operators. On the interior verticals, the product of any two operators gives the third, these objects commuting. On the remaining four lines, the operators anticommute, and the product of any two gives the third with a multiplicative $\pm i$, the plus (minus) depending on the direction of (along/against) the arrow \cite{ref11}.}
\end{figure}

\section{Manipulating density matrices with the elements of the subalgebra}

The Clifford algebra structure and its diagrammatic rendering in Fig. 1 makes operations on the density matrix of $X$-states very simple. Thus, $X_1 \rho X_1^{\dagger}$ leaves $\rho$ in Eq.~(\ref{eqn2}) unchanged. For any of the other six cases, $X_i \rho X_i^{\dagger}$, three coefficients in Eq.~(\ref{eqn2}) remain unchanged (those belonging to that $i$, its conjugate, and 1) while the other four are switched to their negative. As an example of another common operation which occurs, for instance, in the evaluation of ``concurrence" \cite{ref16} or ``quantum discord" \cite{ref2}, $\tilde{\rho} = \sigma_y \tau_y \rho^* \sigma_y \tau_y$ can also be written down from Eq.~(\ref{eqn2}) without any calculation. This involves $X_4$. In Fig. 1, this element is connected to $i=2,3,5,6$ by arrows so that those coefficients $g_i$ have their signs changed by this operation. But the additional complex conjugation involved changes signs again for $i=2,5$ whose $X_i$ in Eq.~(\ref{eqn2}) is pure imaginary. As a result, $\tilde{\rho}$ differs from $\rho$ in Eq.~(\ref{eqn2}) by just flipping the sign of $i=3,6$. In like manner, all such operations become almost automated. And, in evaluating the concurrence through the eigenvalues of $\rho \tilde{\rho}$, with the products of $\{X_i\}$ staying within the algebra and again easily written down with reference to Fig. 1, this construct is again a linear combination of the form of Eq.~(\ref{eqn2}). The square roots of the four eigenvalues turn out to be 

\begin{eqnarray}
\!\!& & \!\!\frac{1}{2}[\sqrt{(1\!+ \! g_1)^2 \! - \! (g_3 \! + \! g_6)^2} \pm\!\! \sqrt{((g_2 \! + \! g_5)^2 \! +\! (g_4 \! + \!g_7)^2}] \nonumber \\
\!\!& & \!\! \frac{1}{2}[\sqrt{(1\! - \! g_1)^2 \! - \! (g_3 \! - \! g_6)^2} \pm\!\! \sqrt{((g_2 \! - \! g_5)^2 \!+\! (g_4 \! -\! g_7)^2}].
\label{eqn4}
\end{eqnarray}
Note the appearance of the combinations of conjugate pairs 2-5, 3-6 and 4-7 of Fig. 1 (and, of course, 1 with the unit element). Arranging the quantities in Eq.~(\ref{eqn4}) in decreasing order and subtracting the sum of the last three from the highest provides the concurrence for any $X$-state.

A prescription for creating $X$-states from a general density matrix can also be easily provided. Each of the fifteen operators of $su(4)$ commutes with six and anticommutes with eight of the others \cite{ref8,ref9}. Thus, $X_1$ which commutes with the six in the $su(2) \times su(2) \times u(1)$ set necessarily anticommutes with all the eight left out of this set. As a result, in a general density matrix $\rho$ of two qubits with all elements nonzero, action by this element, $X_1 \rho X_1^{\dagger}$, changes the signs of the entries in those eight positions which have zeroes in Eq.~(\ref{eqn1}), while leaving other coefficients unchanged. A a result, starting from any general $\rho$, adding to it $X_1 \rho X_1^{\dagger}$ will generate an $X$-state. The action of $X_1(=\sigma_z \tau_z)$ is, of course, the simultaneous unitary rotation through $\pi$ of both spins. 

In the dynamics of two qubits as well, such as in entanglement evolution, including in dissipative processes, the $X$-structure of the states is preserved when the Kraus or other operators involved are expressible in terms of the set $\{X_i\}$.  

\section{Other classes of ``$X$-states"}

The identification above of the ``standard" (as commonly used) $X$-states  with the element $X_1=\sigma_z\tau_z$ immediately points to several other groups of states having the same characteristic of an invariant algebra of $su(2) \times su(2) \times u(1)$ but with different choices for $X_1$. One, for instance, is to choose for this element $\sigma_z$. The other six members are $(\vec{\tau}, \sigma_z \vec{\tau})$. All these matrices and the resulting density matrix that is invariant in form under their operations are now block diagonal $4 \times 4$ matrices, with all eight elements  in the two off-diagonal blocks  zero. While not looking like the X in Eq.~(\ref{eqn1}), they now stand for a decoupling of the 1-2 and 3-4 subspaces, different from that in Eq.~(\ref{eqn1}). In terms of the basis states, they can now be very different in separability and entanglement properties, grouping $(|\uparrow \uparrow 
\rangle, |\uparrow \downarrow \rangle)$ and $ (|\downarrow \uparrow \rangle, 
|\downarrow \downarrow \rangle) $ together instead of $(|\uparrow \uparrow 
\rangle, |\downarrow \downarrow \rangle) $ and $(|\uparrow \downarrow \rangle, |\downarrow \uparrow \rangle)$. Nevertheless, algebraically they are also invariant sets of an $su(2) \times su(2) \times u(1)$ algebra.

Even more interestingly, an $X$-state need not have any zeroes in its density matrix! Thus, the choice, $(X_1 =\sigma_x \tau_x, X_2=\sigma_z \tau_y, X_3=\tau_x, X_4=-\sigma_z\tau_z, X_5=\sigma_y\tau_z, X_6=\sigma_x, X_7=\sigma_y\tau_y)$, built on commuting element $X_1=\sigma_x\tau_x$ is equally valid as an $X$-state with the same invariance algebra, although its density matrix has no zero entries:

\begin{equation}
\rho = \! \frac{1}{4}I \! + \! \frac{1}{4}\left( 
\begin{array}{cccc}
-g_4+g_6 & g_3-ig_2 & -ig_5 & g_1-g_7 \\ 
g_3+ig_2 & g_4+g_6 & g_1+g_7 & ig_5 \\ 
ig_5 & g_1+g_7 & g_4-g_6 & g_3+ig_2 \\
g_1-g_7 & -ig_5 & g_3-ig_2 & -g_4-g_6
\end{array}
\right).      
\label{eqn5}
\end{equation}
Its preparation, by adding to $\rho$ a transformation under $X_1 (=\sigma_x \tau_x)$, and its manipulation or calculation of concurrence and other properties, all proceed as in Section III. Indeed, as can be seen from the choice of the two sets of $\{X_i\}$, it differs from the ``standard" set by a cyclic permutation of the indices $(x, y, z)$ which means two $\pi/2$ rotations of axes, first clockwise with respect to $y$ and then counter-clockwise with respect to an intermediate $z$ axis. Alternatively, this can be represented as the application of $\pi$ pulses to the spins. Clearly, the physics is unchanged with mere change of bases.

Entanglement and other investigations of these other sets will be of interest. A natural extension for further investigation is to analyze similarly sets of invariant states of other subalgebras of $su(4)$ such as $su(3)$ and $so(5)$ \cite{ref10}. The recognition of invariant sets opens a new window into such studies, focusing on what is essential and what are simply changes in bases and representations.

I thank Mr. Mazhar Ali, and Drs. Gernot Alber and Joseph Renes for discussions and for their hospitality at the Technical University, Darmstadt, during this work. This work was supported by the Alexander von Humboldt Foundation.


\begin{thebibliography}{}

\bibitem[*]{} Email: arau@phys.lsu.edu

\bibitem{ref1} See, for instance, M. Nielsen and I. Chuang, {\it Quantum Computation and Quantum Information} (Cambridge University Press, New York, 2000).

\bibitem{ref2} H. Ollivier and W. H. Zurek, Phys. Rev. Lett. {\bf 88}, 017001 (2002); L. Henderson and V. Vedral, {\it ibid} {\bf 90}, 050401 (2003); J. Maziero, L. C. Celeri, R. M. Serra, and V. Vedral, arXiv: 0905.3396.

\bibitem{ref3} T. Yu and J. H. Eberly, Quantum Inform. and Comput. {\bf 7}, 459 (2007); Phys. Rev. Lett. {\bf 93}, 140404 (2004); {\it ibid} {\bf 97}, 140403 (2006).

\bibitem{ref4} See, for instance, S. Bose, I. Fuentes-Guridi, P. L. Knight, and V. Vedral, Phys. Rev. Lett. {\bf 87}, 050401 (2001); G. L, Kamta and A. F. Starace, {\it ibid} {\bf 88}, 107901 (2002); J. S. Pratt, {\it ibid} {\bf 93}, 237205 (2004); S. J. Gu, G. S. Tian, and H. Q. Lin, Phys. Rev. A {\bf 71}, 052322 (2005); J. Wang, H. Batelaan, J. Podany, and A. F. Starace, J. Phys. B {\bf 39}, 4343 (2006).

\bibitem{ref5} R. Dillenschneider, Phys. Rev. B {\bf 78}, 22413 (2008); S. Luo, Phys. Rev. A {\bf 77}, 042303 (2008); M. S. Sarandy, arXiv: 0905.1347; T. Werlang, S. Souza, F. F. Fanchini, and C. J. Villas Boas, arXiv: 0905.3376.

\bibitem{ref6} L. Jakobczyk and A. Jamroz, Phys. Lett. A {\bf 333}, 35 (2004); M. Franca Santos, P. Milman, L. Davidovich, and N. Zagury, Phys. Rev. A {\bf 73}, 040305 (2006); A. Jamroz, J. Phys. A: Math. Gen. {\bf 39}, 727 (2006); M. Ikram, F. L. Li, and M. S. Zubairy, Phys. Rev. A {\bf 75}, 062336 (2007); A. Al-Qasimi and D. F. V. James, {\it ibid} {\bf 77}, 012117 (2007); A. R. P. Rau, M. Ali, and G. Alber, EPL {\bf 82}, 40002 (2008); X. Cao and H. Zheng, Phys. Rev. A {\bf 77}, 022320 (2008); C. E. Lopez, G. Romero, F. Castra, E. Solano, and J. C. Retamal, Phys. Rev. Lett. {\bf 101}, 080503 (2008); M. Ali, G. Alber, and A. R. P. Rau, J. Phys. B: At. Mol. Opt. Phys. {\bf 42}, 025501 (2009). 

\bibitem{ref7} R. F. Werner, Phys. Rev. A {\bf 40}, 4277 (1989).

\bibitem{ref8} A. R. P. Rau, Phys. Rev. A {\bf 61}, 032301 (2000).

\bibitem{ref9} A. R. P. Rau, G. Selvaraj, and D. Uskov, Phys. Rev. A {\bf 71}, 062316 (2005).

\bibitem{ref10} D. Uskov and A. R. P. Rau, Phys. Rev. A {\bf 74}, 030304 (R) (2005); {\it ibid} {\bf 78}, 022331 (2008); Sai Vinjanampathy and A. R. P. Rau, arXiv: 0906.1259.

\bibitem{ref11} A. R. P. Rau, Phys. Rev. A {\bf 79}, 042323 (2009).

\bibitem{ref12} F. J. M. van de Ven and C. W. Hilbers, J. Magn. Reson. {\bf 54}, 512 (1983).

\bibitem{ref13} J. Zhang, J. Vala, S. Sastry, and K. B. Whaley, Phys. Rev. A {\bf 67}, 042313 (2003). 

\bibitem{ref14} T. Beth, D. Jungnickel, and H. Lenz, {\it Design Theory}, Vols. 1 and 2, Encyclopaedia of Mathematics, Vol. 69 (Bibl. Inst., Z\"{u}rich, 1985; Cambridge University Press, 1993).

\bibitem{ref15} See, for instance, G. M. Dixon, {\it Division Algebras: Octonions, Quaternions, Complex Numbers and the Algebraic Design of Physics}, Vol. 290 of Mathematics and its Applications (Kluwer, Dordrecht, 1994); L. E. Dickson, Ann. Of Math. {\bf 20}, 155 (1919); H. S. M. Coxeter, Duke Math. J. {\bf  13}, 561 (1946); J. C. Baez, Bull. Am. Math. Soc. {\bf 39}, 145 (2002) and www.math.ucr.edu/home/baez/octonions/

\bibitem{ref16} S. Hill and W. K. Wootters, Phys. Rev. Lett. {\bf 78}, 5022 (1997); W. K. Wootters, {\it ibid} {\bf 80}, 2245 (1998).

\end{thebibliography}
\end{document}